\documentclass[prx,floatfix,showpacs,superscriptaddress, twocolumn, longbibliography]{revtex4-2}
\usepackage{booktabs}
\usepackage{graphicx}
\usepackage{dcolumn}
\usepackage{bm}
\usepackage{multirow}
\usepackage{amsmath}

\newcommand{\bibcommenthead}[1]{}
\usepackage[colorlinks=true,
linkcolor=blue,
anchorcolor=blue,
urlcolor=blue,
citecolor=blue]{hyperref} 

\begin{document}

\title{Emergent Vortex Ordering in a Multiflavor Pyrochlore-Lattice Compound GeCo$_2$O$_4$}

\author{Jiajun Mo}
\affiliation{Department of Physics, University of Science and Technology of China, Hefei, Anhui 230026,  China}

\author{Otkur Omar}
\affiliation{Department of Physics, University of Science and Technology of China, Hefei, Anhui 230026, China}

\author{Shuangkui Guang}
\affiliation{Department of Physics, University of Science and Technology of China, Hefei, Anhui 230026, China}

\author{Kazuki Iida}
\affiliation{Neutron Science and Technology Centre, Comprehensive Research Organization for Science and Society (CROSS), Tokai, Ibaraki 319-1106, Japan}

\author{Kazuya Kamazawa}
\affiliation{Neutron Science and Technology Centre, Comprehensive Research Organization for Science and Society (CROSS), Tokai, Ibaraki 319-1106, Japan}

\author{Fabio Orlandi}
\affiliation{ISIS Neutron and Muon Source, Rutherford Appleton Laboratory, Didcot OX11 0QX, Oxfordshire UK}

\author{Wenyun Yang}
\affiliation{State Key Laboratory for Mesoscopic Physics, School of Physics, Peking University, Beijing, 100871, China}
\affiliation{Collaborative Innovation Center of Quantum Matter, Beijing, 100871, China}
\affiliation{Beijing Key Laboratory for Magnetoelectric Materials and Devices, Beijing, 100871, China}

\author{Xiaobai Ma}
\affiliation{Department of Nuclear Physics, China Institute of Atomic Energy, Beijing, 102413, China}

\author{Xiquan Zheng}
\affiliation{International Center for Quantum Materials, School of Physics, Peking University, Beijing 100871, China}

\author{Yingying Peng}
\affiliation{International Center for Quantum Materials, School of Physics, Peking University, Beijing 100871, China}

\author{Yuan Xiao}
\affiliation{Department of Physics, University of Science and Technology of China, Hefei, Anhui 230026, China}

\author{Shunhong Zhang}
\affiliation{International Center for Quantum Design of Functional Materials (ICQD), Hefei National Research Center for Physical Sciences at the Microscale, University of Science and Technology of China, Hefei 230026, China}
\affiliation{Hefei National Laboratory, University of Science and Technology of China, Hefei 230088, China}

\author{Oksana Zaharko}
\affiliation{PSI Center for Neutron and Muon Sciences, Villigen, PSI, Switzerland}

\author{Xuefeng Sun}
\email{xfsun@ahu.edu.cn}
\affiliation{Anhui Provincial Key Laboratory of Magnetic Functional Materials and Devices, Institutes of Physical Science and Information Technology, Anhui University, Hefei, Anhui 230601, China}

\author{Shang Gao}%
\email{sgao@ustc.edu.cn}
\affiliation{Department of Physics, University of Science and Technology of China, Hefei, Anhui 230026, China}

\date{\today}
              
\begin{abstract} 
Entangled spin and orbital degrees of freedom provide a multiflavor route to novel magnetic states inaccessible in conventional spin systems. Here, we report the experimental identification of an emergent vortex lattice in the multiflavor pyrochlore-lattice compound GeCo$_2$O$_4$. By combining comprehensive neutron scattering experiments with a regularized regression framework, we identify substantial Kitaev interactions among the nearest-neighboring Co$^{2+}$ pseudospins, which cooperate with geometric frustration to stabilize the vortex order. These results reveal an unexpected route to vortex-lattice order in a three-dimensional Kitaev-frustrated magnet and demonstrate a regularized protocol for Hamiltonian determination in frustrated quantum materials.
\end{abstract}

\maketitle
 
\section{\label{sec:level1} Introduction}

In magnetic materials, exchange interactions are often described by isotropic Heisenberg models~\cite{fazekas_lecture_1999}. However, when couplings to additional degrees of freedom (DOFs) exist, more exotic exchange interactions may emerge, opening a route to novel correlations and applications~\cite{imada_metal_1998, tokura_multiferroics_2014, spaldin_advances_2019}. A well-known ingredient is the orbital DOF: Within the framework of the Kugel-Khomskii model that treats spin and orbital DOFs on equal footing~\cite{kugel_jahn_1982}, entangled spin and orbitals form an extended manifold that is described in a multiflavor basis~\cite{joshi_elementary_1999, chen_multiflavor_2024}. By referring to the lowest-lying manifolds as pseudospins, highly anisotropic couplings like the bond-dependent Kitaev exchange interaction may emerge, a concept that has become a frontier of experimental and theoretical research since its proposal~\cite{takagi_concept_2019, matsuda_kitaev_2025}.

Although the initial quest for Kitaev physics focused on heavy transition metal ions with strong spin-orbit coupling (SOC)~\cite{jackeli_mott_2009}, recent endeavors have expanded towards $3d$ systems~\cite{liu_kitaev_2020, shangguan_one_2023, chen_strength_2024, gu_signatures_2024, lee_fundamental_2020, bandyopadhyay_exchange_2022, songvilay_kitaev_2020, Motome_materials_2020, lin_field_2021, yao_excitations_2022, li_giant_2022}. This shift is not merely about broadening material candidates: As exemplified by the calculations for the Co$^{2+}$ ions~\cite{liu_pseudo_2018, sano_kitaev_2018}, multiflavor $3d$ ions can provide cleaner realizations of Kitaev exchange terms with suppressed isotropic components. In real materials, however, the exact form of the exchange coupling is highly sensitive to the crystal electric fields due to the involvement of $e_g$ electrons~\cite{winter_magnetic_2022,mou_comparative_2024}. This sensitivity leads to a rich variety of exchange interactions among the Co$^{2+}$-based systems, including Kitaev~\cite{zhong_weak_2020, maksimov_strong_2025, xiang_disorder_2023, bischof_spin_2025}, XXZ~\cite{yuan_dirac_2020, sheng_two_2022, xiang_giant_2024}, or Heisenberg~\cite{wildes_spin_2023}  within and beyond the honeycomb lattice geometry~\cite{kim_bond_2023, morris_duality_2021, churchill_transforming_2024, konieczna_understanding_2025}. Therefore, distinguishing genuine Kitaev signatures from competing isotropic interactions remains a priority for the $3d$ candidates.

In this context, while the realization of the Kitaev spin liquid remains elusive~\cite{kitaev_anyons_2006}, exotic magnetic orders such as multi-Q structures have emerged as distinctive signatures for the presence of bond-dependent Kitaev interactions~\cite{kim_bond_2023, kruger_triple_2023}. One prominent example is Na$_2$Co$_2$TeO$_6$~\cite{songvilay_kitaev_2020, lin_field_2021, yao_excitations_2022}, where a Kitaev-driven triple-Q order has been proposed~\cite{kruger_triple_2023, jin_robust_2025}, although a conventional single-Q zig-zag order has also been debated~\cite{zhang_electronic_2023, xiang_disorder_2023, jiao_static_2024, bischof_spin_2025}. Identifying such multi-Q orders is notoriously difficult, necessitating the use of indirect evidence such as the anomalous evolution of the magnetic domain population~\cite{gao_spiral_2017, andriushin_reentrant_2024, jin_robust_2025} or comprehensive microscopic modeling~\cite{gao_fractional_2020, paddison_suppressed_2021, paddison_cubic_2024, mo_skyrmion_2025, andriushin_anomalous_2025}.

Motivated by the potential for multi-Q orders in Kitaev systems, we extend our investigation into the multiflavor pyrochlore-lattice compound GeCo$_2$O$_4$, where DFT calculations have suggested the presence of Kitaev couplings~\cite{liu_pseudo_2018}. While pyrochlores hosting 4$f$ rare-earth ions with anisotropic interactions have been extensively investigated in the context of quantum spin ice~\cite{ross_quantum_2011, gingras_quantum_2014, yan_theory_2017}, the realization of bond-dependent Kitaev exchange in 3$d$ transition-metal pyrochlores remains scarce~\cite{krizan_nccf_2014, ross_static_2016, ross_single_2017, frandsen_real_2017}. To date, studies of anisotropic models in the 3$d$ transition-metal pyrochlores have predominantly focused on the Dzyaloshinskii-Moriya interactions (DMI), as exemplified by Lu$_2$V$_2$O$_7$~\cite{onose_observation_2010, mena_spin_2014}, with only a few exceptions like NaCaNi$_2$F$_7$, where the existence of relatively weak off-diagonal exchange couplings has been established~\cite{plumb_continuum_2019}. To bridge this gap, we combine comprehensive neutron scattering experiments and a methodological innovation of regularized regression based on an effective parameter count. This approach provides strong evidence for substantial bond-dependent Kitaev interactions in GeCo$_2$O$_4$. Moreover, our microscopic modeling reveals the emergence of a novel double-Q vortex lattice composed of entangled spin and orbital pseudospins. This vortex order is experimentally verified via magnetic-field-dependent neutron diffraction signatures, which distinguish it from the previously assumed single-Q collinear order~\cite{diaz_magnetic_2006,matsuda_magnetic_2011,fabreges_field_2017}. 

\begin{figure}[h]
    % \centering
    \hypertarget{anchor:fig1-top}{}
    \includegraphics[width=1.0\linewidth]{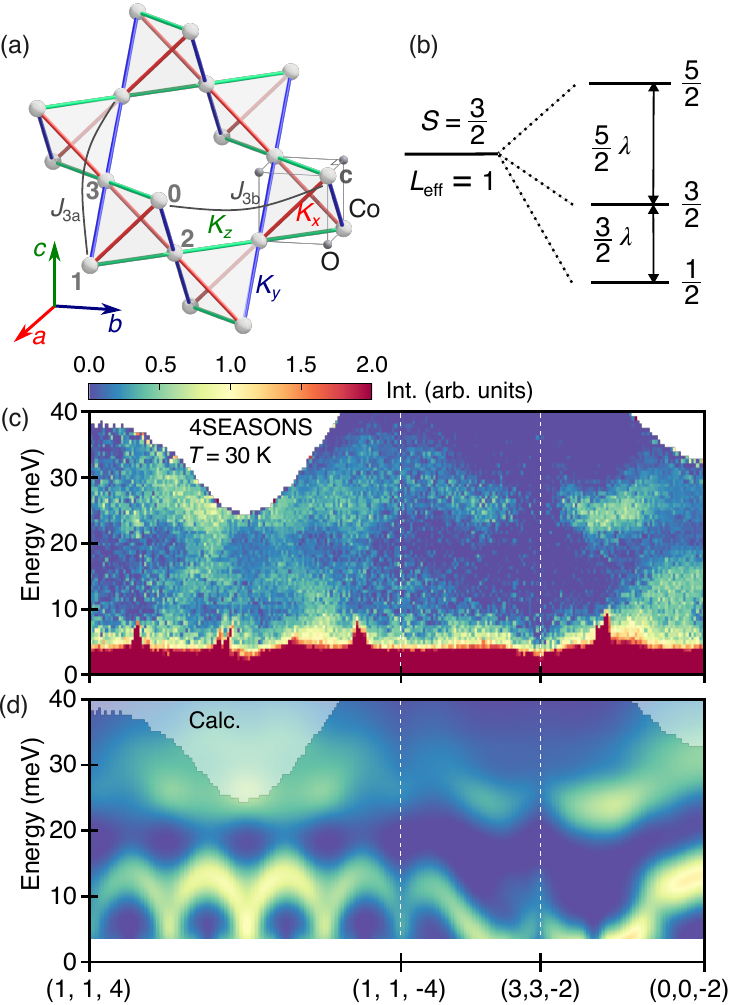}
    \caption{(a) Schematic plot of the pyrochlore lattice formed by the magnetic Co$^{2+}$ ions in GeCo$_2$O$_4$. The first-neighboring bonds, $J_1$, within the $ab$, $bc$, and $ac$ planes are indicated in green, red, and blue, respectively. The exchange pathways for the symmetry-inequivalent $J_\mathrm{3a}$ and $J_\mathrm{3b}$ couplings are also indicated. On the bottom right, positions of the oxygen ions around the tetrahedron are explicitly shown, emphasizing the 90$^\circ$ Co-O-Co exchange paths between the first neighboring Co$^{2+}$. (b) Schematic of the SOC-induced energy level splitting for a high-spin Co$^{2+}$ ion in an octahedral crystal field, with $\lambda$ representing the strength of the SOC. (c) INS spectra, $S(\mathbf Q,\omega)$, for GeCo$_2$O$_4$ measured on the 4SEASONS spectrometer at $T$ = 30~K with an incident energy of $E_i = 60$~meV. (d) Simulated INS spectra for the isotropic $J_1$-$J_\mathrm{3b}$ model using the MF-RPA method, with $J_1 = -1.10$, $J_\mathrm{3b}$ = 0.21~meV.} 
    \label{fig:fig1}
\end{figure}

\section{Paramagnon and spin-orbital excitations}

The pyrochlore lattice formed by the Co$^{2+}$ ions in GeCo$_2$O$_4$ is described in Fig.~\ref{fig:fig1}(a). The first-neighboring bonds, shown in three different colors according to their directions, are mediated through two 90$^\circ$ Co-O-Co exchange paths over edge-sharing CoO$_6$ octahedra, a known geometry favoring the Kitaev interactions~\cite{jackeli_mott_2009,liu_pseudo_2018}. Within each octahedron, the Co$^{2+}$ ion adopts a high-spin state ($|\mathbf{S}| = \frac{3}{2}$) with an unquenched effective orbital angular momentum ($|\mathbf{L}_\mathrm{eff}| = 1$). As schematically illustrated in Fig.~\ref{fig:fig1}(b), the SOC, $\mathcal{H}_\mathrm{SOC} = \lambda\mathbf{L}_{\mathrm{eff}}\cdot\mathbf{S}$, lifts the degeneracy of the spin-orbital manifold, leading to three multiplets characterized by the total effective angular momentum $J_\mathrm{eff} = \frac{1}{2}$, $\frac{3}{2}$, and $\frac{5}{2}$.

 \begin{figure*}[t]
    % \centering
    \hypertarget{anchor:fig2-top}{}
    \includegraphics[width=1.0\linewidth]{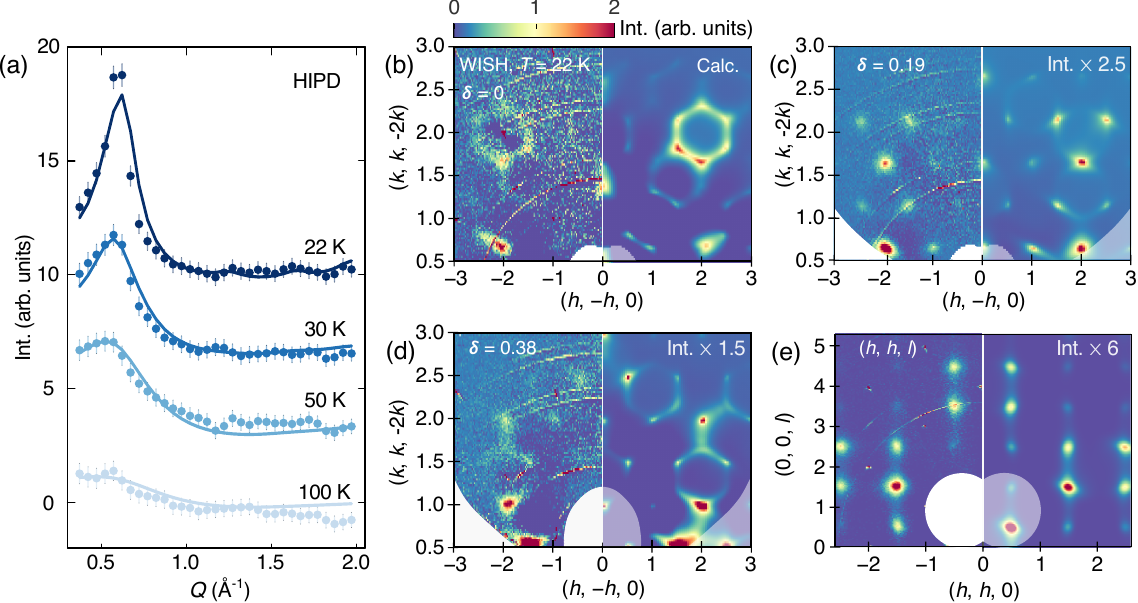}
    \caption{(a) Neutron diffraction pattern for powder GeCo$_2$O$_4$ collected on the HIPD diffractometer at $T$ = 22, 30, 50, and 100~K. Experimental and the corresponding simulated data calculated for the optimal spin-$\frac{1}{2}$ model are shown as circular markers and solid lines, respectively. The patterns at $T = 50$, 30, and 22~K are offset vertically for clarity by 3.5, 7.0, and 10.5 units, respectively. (b-e) Comparison between the experimental and simulated diffuse neutron scattering patterns for crystal GeCo$_2$O$_4$. The experimental data, shown on the left part of each panel, were collected on the WISH diffractometer at $T$ = 22~K. Slices in panels (b-d) are perpendicular to the [111] direction, shifted by $(\delta, \delta, \delta)$ from the Brillouin center with $\delta = 0$ (b), 0.19 (c), and 0.38~r.l.u. (d). Slice in panel (e) was collected in the $(h,h,l)$ plane. The integration width along the vertical direction is 0.1~r.l.u. in all panels.}
    \label{fig:fig2}
\end{figure*} 

Previous inelastic neutron scattering (INS) experiments on powder GeCo$_2$O$_4$ performed at temperatures above $T_N \sim 21$~K observed the existence of two broad excitations at energy transfers of $\sim16$ and 29~meV. While the latter was ascribed to the excitation from the $J_\mathrm{eff} = \frac{1}{2}$ to the $\frac{3}{2}$ manifold, the former was interpreted as local molecular excitations with weak dispersion~\cite{tomiyasu_molecular_2011}. Figure~\ref{fig:fig1}(c) presents our INS spectra on single-crystal GeCo$_2$O$_4$ measured at $T = 30$~K on the 4SEASONS spectrometer at the J-PARC MLF~\cite{kajimoto_fermi_2011}. Despite thermal fluctuations broadening the excitations, two highly dispersive modes with a similar bandwidth of $\sim 15$~meV are resolved, revealing that the spin dynamics in GeCo$_2$O$_4$ are much more collective than previously anticipated~\cite{tomiyasu_molecular_2011}.

Starting from a multiflavor scheme that incorporates the SOC and the crystal electric field, we utilized the mean-field random-phase-approximation (MF-RPA) method to analyze the INS spectra in the paramagnetic regime~\cite{supp}. As compared in Figs.~\ref{fig:fig1}(c) and (d), the INS spectra can be described by a minimal model that considers dominant ferromagnetic first-neighbor coupling, $J_1$, and sub-dominant third-neighbor couplings, $J_\mathrm{3b}$. Since several nearby parameter sets can reproduce the broad paramagnetic response at this temperature~\cite{supp}, this comparison is intended to establish the collective character and approximate energy scale of the excitations, rather than to provide an independent high-precision determination of the exchange parameters. Within this scope, the good agreement between the experimental and calculated INS spectra confirms the high-energy branch as a spin-orbital excitation and demonstrates that the spin dynamics associated with the low-energy branch at $E \lesssim 16$~meV are more appropriately described as paramagnons rather than molecular modes. 

\begin{figure*}[t]
    % \centering
    \hypertarget{anchor:fig3-top}{}
    \includegraphics[width=1.0\linewidth]{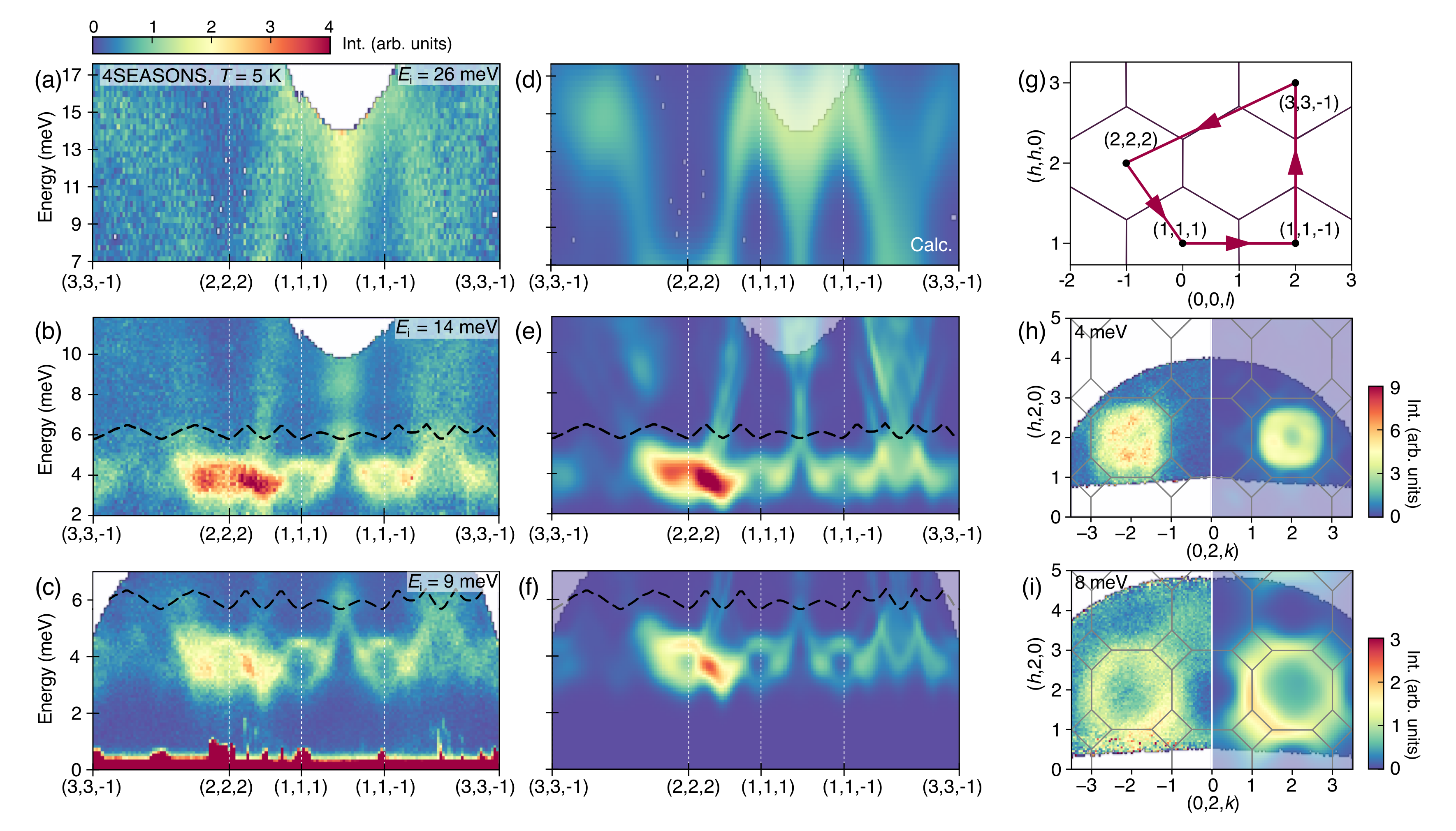}
    \caption{(a-c) Single-crystal INS spectra for GeCo$_2$O$_4$ measured at $T = 5$~K on the 4SEASONS spectrometer with an incident energy of $E_\mathrm i$  = 26.0 (a), 14.4 (b), and 9.2~meV (c). (d-f) Corresponding INS spectra calculated using the LSWT for the fitted model. (g) The high-symmetry path in reciprocal space for the spectra shown in (a)-(f). In panels (b-f), the lower boundary of the two-magnon continuum calculated using the \texttt{SpinToolkit} program~\cite{sptk,XuL2026} is indicated by the dashed line. (h, i) The left part presents the constant-energy slice of the experimental INS spectra in the ($h$, $h$, $l$) plane at energy transfer of 4 (h) and 8~meV (i). The right part is calculated for the fitted model using the LSWT.}
    \label{fig:fig3}
\end{figure*} 

\section{Revisiting the molecular modes}

To further test the validity of the molecular models~\cite{tomiyasu_molecular_2011} and provide additional constraints for microscopic modeling~\cite{bai_hybridized_2021, gao_spiral_2022}, we performed powder and single-crystal neutron diffraction experiments on GeCo$_2$O$_4$ to clarify its quasielastic spin correlations in the paramagnetic regime. Powder neutron diffraction provides a highly efficient probe for the short-range spin correlations as a function of temperature. Figure~\ref{fig:fig2}(a) presents the powder neutron diffraction data for GeCo$_2$O$_4$ collected on the HIPD diffractometer at the China Advanced Research Reactor (CARR)~\cite{supp}. At $T = 22$~K, slightly above $T_\mathrm N$, a broad peak around wavevector transfer $Q = 0.6\text{\AA}^{-1}$ is observed, which corresponds to short-range spin fluctuations with magnetic propagation vector $\bm{q} = (\frac{1}{2}, \frac{1}{2}, \frac{1}{2})$~\cite{matsuda_magnetic_2011,diaz_magnetic_2006}. As expected for a $J_1$-$J_3$ pyrochlore-lattice model with dominant ferromagnetic $J_1$ interactions~\cite{gao_line_2022}, short-range spin correlations survive at temperatures up to 100~K, which aligns with the Weiss temperature of $\Theta \sim 81$~K fitted from the magnetic susceptibility measurements~\cite{diazStudy2004a,lashley_specific_2008}.

In Figs.~\ref{fig:fig2}(b-e), the left halves present the single-crystal diffuse scattering patterns for GeCo$_2$O$_4$ measured on the WISH diffractometer at the ISIS spallation neutron source~\cite{supp}. In the [111] slice shown in Fig.~\ref{fig:fig2}(b), the diffuse signal forms a hexagonal ring, connecting the tails of the incipient magnetic Bragg peaks that are better resolved in Figs.~\ref{fig:fig2}(c) and (d). In the ($h$,$h$,$l$) slice shown in panel (e), intensity appears as strong diffuse spots near the magnetic Bragg peak positions, coexisting with weak diffuse rods along the (0,0,$l$) direction. These patterns deviate from the expected diffuse signal of the previously assumed molecules~\cite{tomiyasu_molecular_2011}, indicating that the short-range spin fluctuations in GeCo$_2$O$_4$ are not approximated by molecular correlations. Nevertheless, the sharp features revealed by single-crystal diffuse neutron scattering strongly suggest that the microscopic spin Hamiltonian in  GeCo$_2$O$_4$ should go beyond a simple Heisenberg model~\cite{gao_line_2022}.

\section{Regularized Modeling Identifies Nearest-Neighbor Kitaev Couplings}

Further insights into the spin dynamics can be drawn from the INS spectra measured in the long-range ordered regime, where a thorough mapping of the dispersion remains lacking in previous studies~\cite{lashley_specific_2008, tomiyasu_molecular_2011}. Figures~\ref{fig:fig3}(a)-\ref{fig:fig3}(c) present our INS spectra for GeCo$_2$O$_4$~collected on 4SEASONS at $T = 5$~K. Data collected at various incident energies $E_i$ are shown along the high-symmetry directions defined in Fig.~\ref{fig:fig3}(g). In the spectra collected with $E_i = 26.0$~meV shown in Fig.~\ref{fig:fig3}(a), steep dispersive modes up to $\sim 17$~meV are observed, which correspond to the paramagnon branch observed in Fig.~\ref{fig:fig1}. At lower $E_i$ of 14.4 (b) and 9.2~meV (c), an excitation gap of $\sim3$~meV is resolved, indicative of Ising-type anisotropy in this system. Notably, the constant-$E$ slices presented in Figs.~\ref{fig:fig3}(h) and \ref{fig:fig3}(i) at energy transfers of $8$ and $4$~meV reveal a pattern resembling that of the previously proposed molecular states~\cite{tomiyasu_molecular_2011}. This suggests that such molecules may serve as phenomenological descriptions of spin excitations emerging from a more fundamental lattice model, consistent with recent findings for MgCr$_2$O$_4$ and LiGaCr$_4$O$_8$~\cite{gao_dynamic_2024}.

To establish the microscopic model for GeCo$_2$O$_4$, we first resort to the effective spin-$\frac{1}{2}$ scenario by integrating out the spin-orbital excitations centered at 29~meV.  Following the analysis of the INS spectra in the paramagnetic regime as presented in Fig.~\ref{fig:fig1}, we limit the anisotropic terms to the main components of the $J_1$ and $J_\mathrm{3b}$ exchanges. Specifically, for the first-neighboring $\langle12\rangle$ bond and the third-neighboring $\langle 0\mathrm{c} \rangle$ bond indicated in Fig.~\ref{fig:fig1}(a), their generic exchange matrices are expressed as

\begin{equation}
\mathbf{J}^{\langle12\rangle}_{1} = 
\begin{pmatrix}
\phantom{-}J_{1}^{\mathrm A} & {-}J_{1}^{\mathrm C} & \phantom{-}J_{1}^{\mathrm D} \\[4pt]
{-}J_{1}^{\mathrm C} & \phantom{-}J_{1}^{\mathrm A} & {-}J_{1}^{\mathrm D} \\[4pt]
{-}J_{1}^{\mathrm D} & \phantom{-}J_{1}^{\mathrm D} & \phantom{-}J_{1}^{\mathrm B}
\end{pmatrix},
\label{eq:J12_global_ABCD}
\end{equation}

\begin{equation}
\mathbf{J}^{\langle0\mathrm{c}\rangle}_{\mathrm{3b}} = 
\begin{pmatrix}
J_{\mathrm{3b}}^{\mathrm A} & \phantom{-}J_{\mathrm{3b}}^{\mathrm D} & \phantom{-}J_{\mathrm{3b}}^{\mathrm C} \\[4pt]
J_{\mathrm{3b}}^{\mathrm D} & \phantom{-}J_{\mathrm{3b}}^{\mathrm A} & \phantom{-}J_{\mathrm{3b}}^{\mathrm C} \\[4pt]
J_{\mathrm{3b}}^{\mathrm C} & \phantom{-}J_{\mathrm{3b}}^{\mathrm C} & \phantom{-}J_{\mathrm{3b}}^{\mathrm B}
\end{pmatrix}.
\label{eq:J3b_global_ABCD}
\end{equation}
The coupling matrices for other symmetry-related bonds are listed in the Supplemental Materials~\cite{supp}. For each exchange matrix, the difference between the diagonal elements, $J^\mathrm B - J^\mathrm A$, corresponds to the strength of the bond-dependent Kitaev interactions, $J^\mathrm K$~\cite{kimchi_kitaev_2014}, while the off-diagonal elements $J^\mathrm{C}$ and $J^\mathrm{D}$ represent the remaining symmetry-allowed anisotropic exchanges. In particular, the antisymmetric part of the first-neighbor matrix, parameterized by $J^\mathrm{D}_1$, corresponds to the DMI. Such antisymmetric DMI terms are symmetry-forbidden on the third-neighbor $J_\mathrm{3b}$ bonds, for which $J^\mathrm{C}_\mathrm{3b}$ and $J^\mathrm{D}_\mathrm{3b}$ denote symmetric off-diagonal anisotropies.

\begin{figure}[t]
    \hypertarget{anchor:fig-reg}{}
    \includegraphics[width=1.0\linewidth]{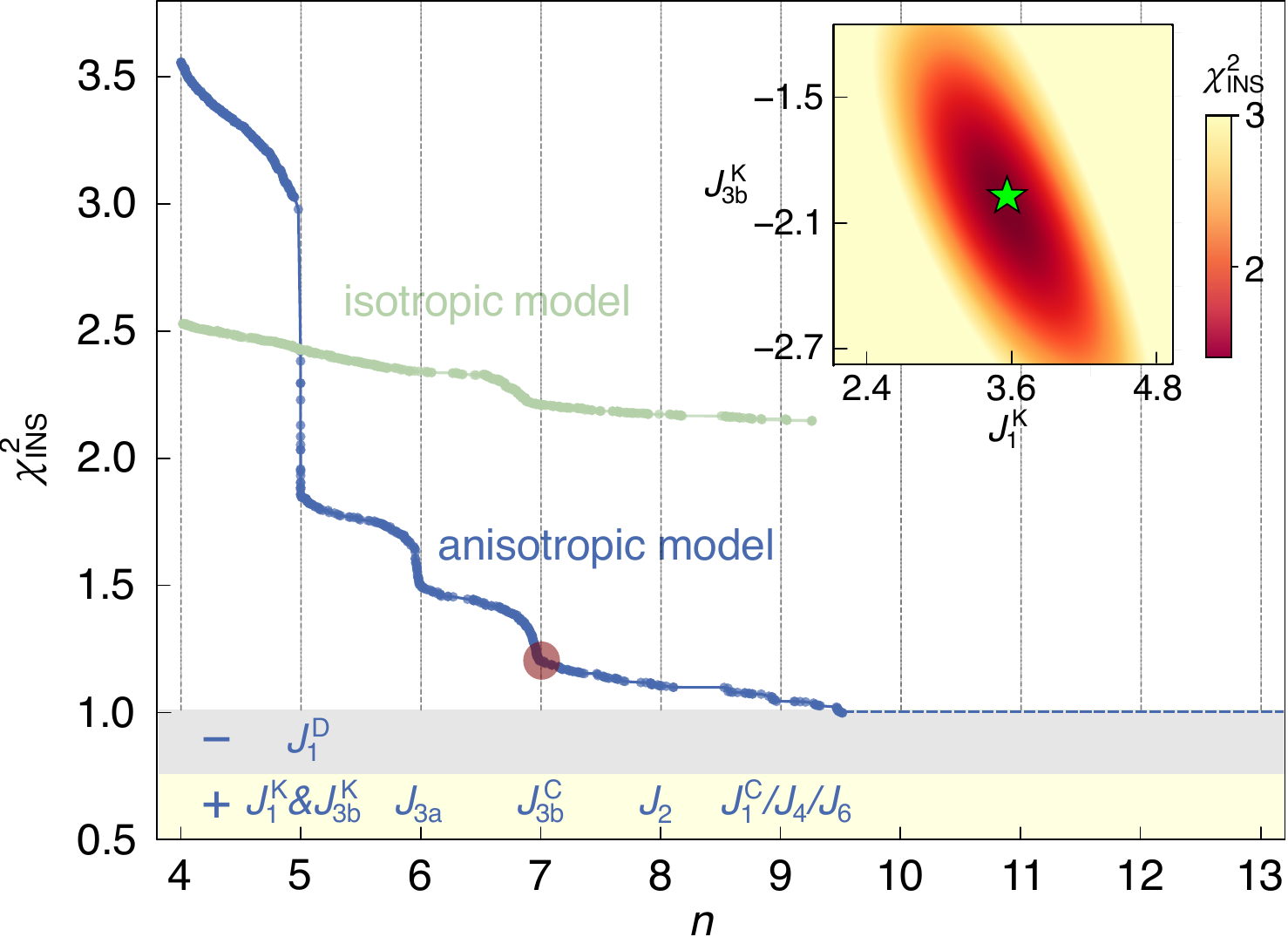}
    \caption{Pareto front obtained from two-target fitting for the anisotropic and isotropic models. Red circle at $n=7$ marks the position of the minimal model. For each integer $n$, the labels at the bottom indicate the parameter changes relative to $n-1$, where the removed and added parameters are highlighted with a gray and yellow background, respectively. At $n=9$, either $J_1^\mathrm{C}$, $J_4$, or $J_6$ can be added to the optimal parameter set. Further addition to the parameter set at $n>9$ does not improve $\chi^2_\mathrm{INS}$ as indicated by the dashed horizontal line. The initial parameter set at $n=4$ is $J_{1}^\mathrm A$, $J_1^\mathrm{D}$, $J_\mathrm{3b}^\mathrm A$ and $J_\mathrm{3b}^\mathrm D$. Inset shows the $\chi^2$ mapping as a function of $J_1^\mathrm{K}$ and $J_\mathrm{3b}^\mathrm{K}$. Position of the optimal parameter set was indicated by the green star marker.}
    \label{fig:fig-reg}
\end{figure}

While conventional model refinement employs least-squares fitting, its application to complex frustrated systems risks overfitting in expanded parameter spaces. Consequently, fitting results become highly sensitive to manual parameter selection. To address this, we adapt the regularized regression technique~\cite{Louizos2018Learning,barron2019general} to establish the minimal model for GeCo$_2$O$_4$ through an innovative two-target fitting approach. One target, shown along the $y$ axis of Fig.~\ref{fig:fig-reg}, is the goodness-of-fit for the INS spectra, $\chi^2_\mathrm{INS}$, where the theoretical spectra are calculated using linear spin wave theory (SUNNY program~\cite{dahlbomSunnyjl2025}). The other target, shown along the $x$ axis, is an effective parameter count defined as $n = \max[4.0, \sum_i p_i^2 / (p_i^2 + \epsilon^2)]$, where $p_i$ for the anisotropic model describes the 13 parameters that define the exchange couplings up to sixth neighbors, including ($J^\mathrm{A}_1$, $J^\mathrm{K}_1$, $J^\mathrm{C}_1$, $J^\mathrm{D}_1$), $J_2$, $J_\mathrm{3a}$, ($J_\mathrm{3b}^\mathrm{A}$, $J_\mathrm{3b}^\mathrm{K}$, $J_\mathrm{3b}^\mathrm{C}$, $J_\mathrm{3b}^\mathrm{D}$), $J_4$, $J_5$, and $J_6$ in units of meV. The threshold parameter $\epsilon = 0.02$~meV automatically suppresses
marginal parameters without manual intervention. From these two-target fits, the Pareto front, shown in Fig.~\ref{fig:fig-reg}, represents the  optimal INS fits achievable for different parameter counts. The steepness of the Pareto front across integer parameter counts reveals the relative importance of each added parameter in the anisotropic model, shown at the bottom. 

As compared in Fig.~\ref{fig:fig-reg}, the Pareto front of the anisotropic model (blue) is much lower than that of an isotropic dipole model (green) with local easy-axis anisotropy and isotropic couplings up to the 8th neighbors. From the knee points over the Pareto front, the minimal model for GeCo$_2$O$_4$ requires 7 parameters as indicated by the red circular marker in Fig.~\ref{fig:fig-reg}, yielding the Hamiltonian:
\begin{equation}
\mathcal{H} = \sum_{\langle ij\rangle\in 1}J_1^{uv}\mathbf S_i^u·\mathbf S_j^v+\sum_{\langle ij\rangle \in 3a}J_\mathrm{3a}\mathbf S_i·\mathbf S_j+\sum_{\langle ij\rangle\in 3b}J_\mathrm{3b}^{uv}\mathbf S_i^u·\mathbf S_j^v\mathrm{ .}
\end{equation}
The fitted coupling strengths are listed in Table~\ref{tab:exchange}. The smallest retained coupling, $J_\mathrm{min}=J_\mathrm{3a} = 0.47$~meV, is more than one order of magnitude larger than the threshold $\epsilon=0.02$~meV. This separation of scales shows that $\epsilon$ acts only as a regularization scale for counting marginal near-zero coefficients, rather than as a physical cutoff for the exchange interactions. Therefore, as long as $\epsilon \ll J_\mathrm{min}$, variations in $\epsilon$ do not affect the dominant exchange hierarchy in the fitted Hamiltonian.

As compared in Figs.~\ref{fig:fig2} and \ref{fig:fig3}, this parameter set reproduces both the diffuse scattering patterns in the paramagnetic regime and the INS spectra in the ordered regime. While the global fit is excellent, reproducing the $E_i = 26.0$~meV spectra shown in Fig.~\ref{fig:fig3}(a) requires an extra broadening that is 2.3 times of the theoretical instrumental resolution~\cite{supp}, suggesting possible magnon decay when overlapping with the two-magnon continuum~\cite{stoneQuasiparticle2006, zhitomirskyColloquium2013, kim_bond_2023}. The calculated INS spectra for other optimal models with different numbers of effective parameters are presented in the Supplemental Materials~\cite{supp}, showing that models with $n > 7$ do not significantly improve the fit, thus justifying the $n=7$ model as the parsimonious choice as a minimal model. The significant magnitudes of $J^\mathrm K_1$ and $J^\mathrm K_\mathrm{3b}$, which are robust across the Pareto front for $n\ge7$ with standard deviations of 0.18 and 0.06~meV, respectively, reveal substantial bond-dependent Kitaev couplings in the effective spin-$\frac{1}{2}$ basis. This observation is further confirmed by the $\chi^2$ mapping in the inset of Fig.~\ref{fig:fig-reg}, where deviation from the optimal $J_1^\mathrm{K}$ and $J_\mathrm{3b}^\mathrm{K}$ values significantly degrades the fit. 

\begin{table*}[t]
    \setlength{\tabcolsep}{8pt}
    % \centering
    \caption{Fitted exchange parameters in units of meV for the effective spin-$\frac{1}{2}$ model and the multiflavor model with a spin size of $S_{\mathrm{mf}}=\frac{3}{2}$. For the multiflavor model, the fitted parameters are scaled by a factor of $3$ according to the ratio of its spin size compared to that of the effective spin-$\frac{1}{2}$ model.}
    \label{tab:exchange}
    \begin{tabular}{cccccccc}
    \toprule
    \multicolumn{1}{c}{\bfseries Model Type} & 
    \multicolumn{7}{c}{\bfseries Parameters (meV)} \\ 
    \cmidrule(l{3pt}r{3pt}){1-8} 
    % \cmidrule(l{3pt}r{3pt}){2-8}
    \multirow{2}{*}{spin-$\frac{1}{2}$ model}& $J_1^\mathrm A$ & $J_1^\mathrm K$  & $J_\mathrm {3a}$ & $J_\mathrm {3b}^\mathrm A$ & $J_\mathrm {3b}^\mathrm K$ &$J_\mathrm {3b}^\mathrm C$& $J_\mathrm {3b}^\mathrm D$\\
    \cmidrule(l{3pt}r{3pt}){2-8}
     & $-5.60 (7)$ & $3.54 (17)$ & $0.47 (1)$ & $1.81 (3)$  &$-1.97 (7)$ & $0.48 (3)$ & $1.98 (6)$\\
    \midrule
    \addlinespace[2mm] 
    
    \multirow{2}{*}{multiflavor model} & 
    $3J_1^\mathrm A$ & $3J_1^\mathrm K$ & $3J_\mathrm{3a}$ & $3J_\mathrm{3b}^\mathrm A$ & 
    $3J_\mathrm{3b}^\mathrm K$ & $3J_\mathrm{3b}^\mathrm C$ & $3J_\mathrm{3b}^\mathrm D$ \\
    \cmidrule(l{3pt}r{3pt}){2-8}&
    $-5.61(9)$ & $3.27(30)$ &  $0.48(1)$ & $1.29(1)$ &$-$ &$-0.39(2)$ &$0.24(4)$\\
    \bottomrule
    \end{tabular}
\end{table*}

In an effective spin-$\frac{1}{2}$ basis, anisotropic interactions may trivially arise from the projection of single-ion anisotropy~\cite{lines_magnetic_1963}, which motivates us to perform further calculations using a multiflavor Hamiltonian that explicitly incorporates the single-ion anisotropy induced by the crystal electric field~\cite{supp}. Based on the MF-RPA method, we performed least-squares fits for the INS spectra using a minimal model that considers the same 7 exchange parameters as those in the $S=\frac{1}{2}$ model. As compared in Table~\ref{tab:exchange}, the fitted parameter set for the multiflavor model is largely consistent with those of the spin-$\frac{1}{2}$ model, except that the $J_\mathrm{3b}$ term becomes essentially isotropic. The persistence of the $J_1$ anisotropy in the multiflavor description indicates orbital-dependent hopping interference~\cite{liu_pseudo_2018, sano_kitaev_2018}, distinguishing it from the $J_\mathrm{3b}$ interactions, which mainly arise from local single-ion  anisotropy projected onto the effective spin-$\frac{1}{2}$ basis.

\begin{figure*}[t!]
    % \centering
    \hypertarget{anchor:fig4-top}{}
    \includegraphics[width=0.9\linewidth]{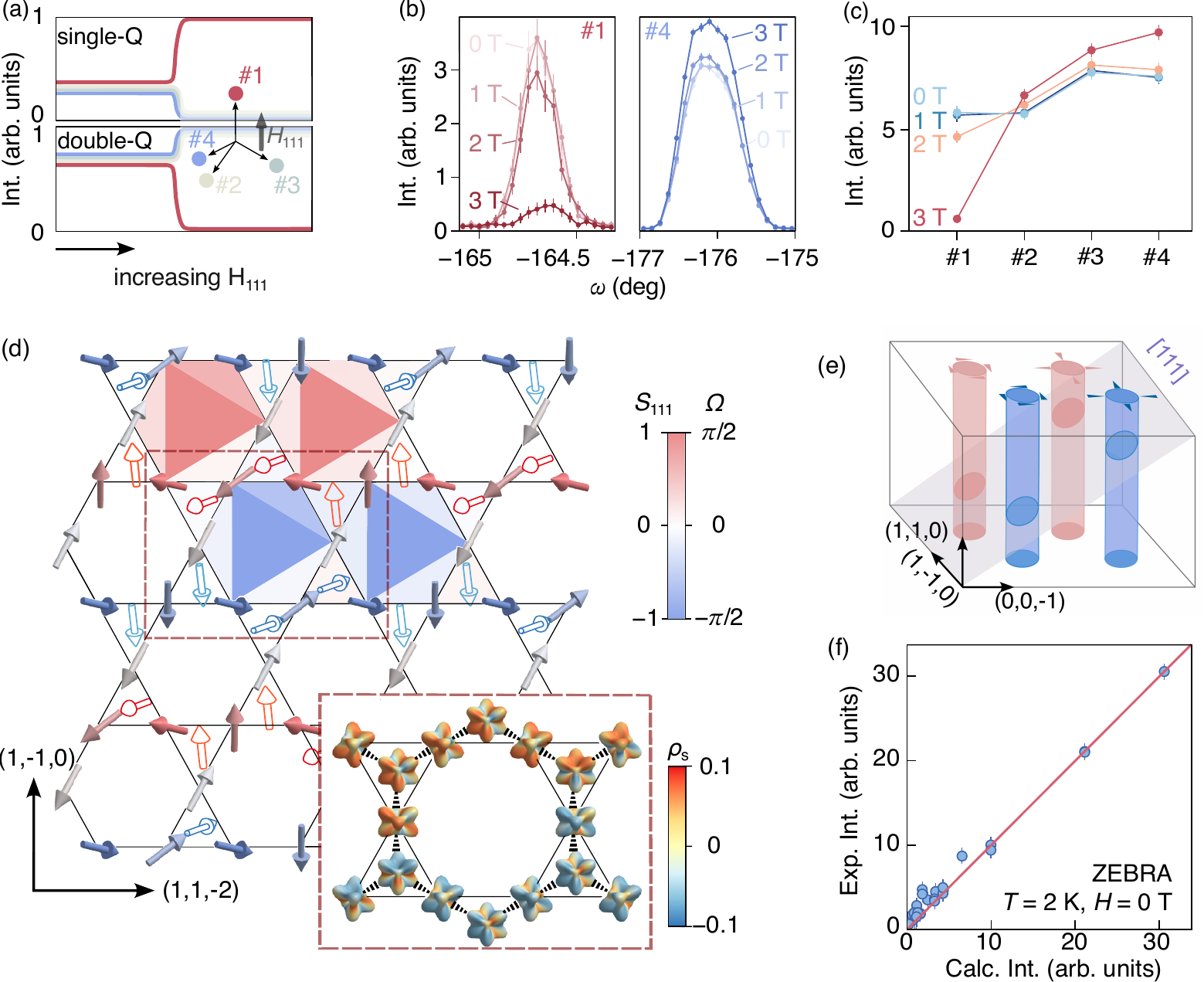}
    \caption{(a) Schematic response of the $\{\frac{1}{2}, \frac{1}{2}, \frac{1}{2}\}$ arms to a magnetic field applied along the [111] direction for the single-Q  (top)  and the double-Q (bottom) structure. The labels $\#1$$\sim$4 correspond to the arms of $\mathbf{q}_1=(\frac{1}{2}, \frac{1}{2}, \frac{1}{2})$, $\mathbf{q}_2=(\frac{1}{2}, \frac{1}{2}, -\frac{1}{2})$, $\mathbf{q}_3=(\frac{1}{2}, -\frac{1}{2}, \frac{1}{2})$, and $\mathbf{q}_4=(-\frac{1}{2}, \frac{1}{2}, \frac{1}{2})$, respectively. (b) Peak profiles of the $\mathbf{Q}_1=(-\frac{5}{2}, \frac{3}{2}, \frac{3}{2})$ and $\mathbf{Q}_4=(-\frac{1}{2}, \frac{1}{2}, \frac{1}{2})$ reflections, which belong to the $\bm{q}_1$ and $\bm{q}_4$ arms, respectively. (c) Field dependence of the integrated intensities for $\mathbf{Q}_1=(-\frac{5}{2}, \frac{3}{2}, \frac{3}{2})$, $\mathbf{Q}_2=(\frac{1}{2}, \frac{1}{2}, -\frac{1}{2})$, $\mathbf{Q}_3=(\frac{1}{2}, -\frac{1}{2}, \frac{1}{2})$, and $\mathbf{Q}_4=(-\frac{1}{2}, \frac{1}{2}, \frac{1}{2})$, which belong to $\bm{q}_1$, $\bm{q}_2$, $\bm{q}_3$, and $\bm{q}_4$ arms, respectively. (d) The vortex lattice formed by pseudospins viewed along the [111] direction. Solid and open arrows represent moments in the kagome and adjacent triangular layers, respectively. Inset presents the corresponding orbital configuration for the area outlined by the dashed rectangle. The colored background represents the calculated topological charge density distribution, which corresponds to the solid angle spanned by the three pseudospins over each triangle.
    (e) Schematics for the vortex lattice, showing the vortex tubes oriented along the [110] direction, penetrating the [111] slice. (f) Comparison of calculated and observed intensities of the magnetic Bragg peaks, with a goodness-of-fit parameter of $R = 12.4\,\%$. The experimental dataset was collected on ZEBRA at $T = 2$~K in zero field.}
    \label{fig:fig4}
\end{figure*} 

\section{Emergent vortex lattice of pseudospins}

Beyond elucidating the spin dynamics, our microscopic modeling of GeCo$_2$O$_4$ also reveals its unique magnetic ground state. Structure factor calculations, shown in the Supplemental Materials~\cite{supp}, reveal a double-Q character for the ground state, which challenges the previously established single-Q order with collinear alignment~\cite{diaz_magnetic_2006,matsuda_magnetic_2011,fabreges_field_2017}. As discussed in the Supplemental Materials~\cite{supp}, the double-Q structure produces magnetic Bragg peaks only at the constituent single-Q propagation-vector positions, without additional multi-Q Bragg peaks. Consequently, conventional refinement of the zero-field magnetic Bragg intensities cannot uniquely distinguish a double-Q state from equally populated single-Q domains.

To address this ambiguity, we performed neutron diffraction experiments on the ZEBRA diffractometer at the Paul Scherrer Institut (PSI) with a magnetic field along [111]~\cite{supp}. In the assumed collinear single-Q order, spins are perpendicular to the propagation vectors~\cite{fabreges_field_2017}; thus, the applied field should stabilize the $\bm{q}_1=  (\frac{1}{2}, \frac{1}{2}, \frac{1}{2})$ arm while suppressing the other three, as illustrated in the top panel of Fig.~\ref{fig:fig4}(a). In contrast, for the double-Q order, our simulations predict an opposite field dependence, with domains containing the $\bm{q}_1$ arm suppressed by the applied field. Figure~\ref{fig:fig4}(b) compares the sample rotation scans for the $\mathbf{Q}_1=(-\frac{5}{2}, \frac{3}{2}, \frac{3}{2})$ and $\mathbf{Q}_4=(-\frac{1}{2}, \frac{1}{2}, \frac{1}{2})$ reflections, which belong to the $\bm{q}_1$ and $\bm{q}_4$ arms, respectively. Measurements were performed at $T = 2$~K by ramping up the field to the designated strengths up to 3~T, all below the meta-magnetic transition reported at $4.2$~T~\cite{hoshiMagnetic2007a}. The suppressed $\bm{q}_1$ arm and enhanced $\bm{q}_4$ arm strongly support the double-Q scenario. Figure~\ref{fig:fig4}(c) summarizes the evolution of the integrated intensities for the characteristic reflections belonging to each of the four arms. The similar field enhancement observed for the $\bm{q}_2$, $\bm{q}_3$, and $\bm{q}_4$ arms further confirms the double-Q scenario as illustrated in Fig.~\ref{fig:fig4}(a).

The magnetic structure of the zero-field ground state, viewed along the [111] direction, is plotted in Fig.~\ref{fig:fig4}(d) for a representative double-Q domain formed by the superposition of $\bm{q}_2$ = ($\frac{1}{2}$, $\frac{1}{2}$, $-\frac{1}{2}$) and $\bm{q}_3=  (\frac{1}{2}, -\frac{1}{2}, \frac{1}{2})$. Ordered moments within a kagome layer and its two neighboring triangular layers are indicated by solid and open arrows, respectively. This double-Q order, described by the $C_a2/c$ magnetic space group~\cite{supp}, emerges as the zero-field magnetic ground state of the fitted Hamiltonian, and is consistent with neutron diffraction experiments on single crystals [Fig.~\ref{fig:fig4}(f)] and powder samples (Supplemental Materials~\cite{supp}). The magnitude of the ordered moment is fitted to be 3.12(5)~$\mu_\mathrm{B}$, consistent with the expected moment size of the Co$^{2+}$ ions~\cite{diaz_magnetic_2006,matsuda_magnetic_2011,fabreges_field_2017}. As evidenced by the magnetic structure in Fig.~\ref{fig:fig4}(d) and the topological charge density defined as the solid angle spanned by the three pseudospins over each triangle~\cite{supp}, the double-Q phase manifests as an emergent vortex lattice. We use the term vortex lattice to denote the periodic array of circulating transverse pseudospin textures in the magnetic unit cell. For each elementary plaquette normal to [111], we characterize the circulation by the discrete winding of the in-plane pseudospin angle around the plaquette. The resulting configuration forms alternating vortex and antivortex tubes, so that the net winding over the magnetic unit cell vanishes while the local vorticity remains finite. Distinct from typical topological magnetic textures where the vortex tubes are perpendicular to the plane~\cite{nagaosa_topo_2013}, the superposition of $\bm{q}_2$ and $\bm{q}_3$ results in vortex tubes oriented along the non-orthogonal [110] direction, which is schematically depicted in Fig.~\ref{fig:fig4}(e). 

Crucially, the multiflavor character of the Co$^{2+}$ ions implies that this vortex lattice involves entangled spin and orbital degrees of freedom. In the region outlined by the dashed rectangle in Fig.~\ref{fig:fig4}(d), the corresponding orbital configuration is detailed in the inset, where the orbital distortion along the local [111] axes is exaggerated for clarity. The color mapping represents the local spin projection density, $\rho_s$, defined along the ordered moment direction at each Co$^{2+}$ site. Because the orbital configuration couples directly to the lattice through the crystal electric field, the magnetic ordering in GeCo$_2$O$_4$ may induce the weak tetragonal distortion observed by synchrotron x-ray diffraction~\cite{barton_structural_2014, fabreges_field_2017}. In the Supplemental Materials~\cite{supp}, we construct the symmetry-allowed Landau free-energy invariants and demonstrate that a bilinear coupling term exists between the double-Q magnetic order parameter and the $\Gamma_{3}^+$ strain associated with the tetragonal distortion. This confirms that the observed tetragonal distortion is  compatible with the vortex-lattice state, although its existence is not decisive for distinguishing the single-Q and double-Q orders~\cite{supp}.  

As the tetragonal strain is very small, $\Delta a/a\sim 1.2\times10^{-3}$~\cite{barton_structural_2014, fabreges_field_2017}, and is below the resolution of our laboratory single-crystal XRD measurements~\cite{supp}, its leading effect is therefore expected to be a weak splitting of symmetry-equivalent cubic domains and exchange paths, rather than a change of the dominant exchange hierarchy in our fitted Hamiltonian. To test whether such a weak lowering of cubic symmetry could alter our conclusions, we introduced tetragonal-symmetry-preserving splittings of the exchange parameters, with relative amplitudes up to $\delta= 0.5\%$ that exceeds the relative lattice distortion scale, re-optimized the classical ground state, and recalculated the field-dependent magnetic Bragg intensities. Over this range, the double-Q ground state and the characteristic field-dependent intensity pattern remain unchanged, thus verifying the robustness of the vortex lattice against weak tetragonal distortion.

\section{Discussion}

The discovery of a vortex lattice in GeCo$_2$O$_4$ establishes a novel topological magnetic texture on the pyrochlore lattice. Notably, the whirling of the ordered moments in GeCo$_2$O$_4$ occurs on an atomic scale of merely $\sim 6$~\AA, nearly two orders of magnitude smaller than known topological spin textures on pyrochlore lattices~\cite{seki_observation_2012, kezsmarki_neel_2015}. This short periodicity, determined by the propagation vector $\bm{q}$ = ($\frac{1}{2}$, $\frac{1}{2}$, $\frac{1}{2}$), is stabilized by competing $J_1$ and $J_3$ couplings~\cite{raju_transition_1999}. As observed in Gd$_2$Ti$_2$O$_7$~\cite{paddison_suppressed_2021, raju_transition_1999, javanparast_fluctuation_2015}, the unique $\bm{q}$ = ($\frac{1}{2}$, $\frac{1}{2}$, $\frac{1}{2}$) implies unequal ordered moments within a single-Q structure, suggesting the natural tendency toward multi-Q orders in pyrochlore systems hosting similar types of propagation vectors~\cite{paddison_suppressed_2021, lee_neel_2008, basu_magnetic_2020, chaix_from_2026}.

While geometric frustration inherent to the pyrochlore lattice drives the double-Q order, the specific vortex lattice demonstrated in GeCo$_2$O$_4$ is selected by the experimentally fitted Hamiltonian containing substantial nearest-neighbor Kitaev anisotropy. Unlike the pyrochlores~\cite{ross_quantum_2011, gingras_quantum_2014, yan_theory_2017}, where magnetic rare-earth ions reside in the 8-fold oxygen coordinated sites, GeCo$_2$O$_4$ realizes this anisotropic exchange through Co$^{2+}$ ions in edge-sharing octahedral environments, providing a distinct platform to test Kitaev physics in three-dimensional networks. Although GeCo$_2$O$_4$ exhibits long-range magnetic order, the substantial Kitaev anisotropy identified here suggests that field- or pressure-tuned studies could provide a useful route to explore nearby competing phases.

Methodologically, our work demonstrates that Pareto front analysis provides a systematic regularized criterion for model selection in frustrated magnets, reducing the subjectivity associated with manual parameter pruning. This protocol is particularly valuable for multiflavor systems where anisotropic interactions multiply the number of exchange components. Pronounced examples include RuCl$_3$~\cite{wu_field_2018, ran_spin_2017, banerjee_proximity_2016, ozel_magnetic_2019}, VI$_3$~\cite{gu_signatures_2024, shen_magnetoelastic_2026}, Na$_2$Co$_2$TeO$_6$~\cite{songvilay_kitaev_2020, lin_field_2021, yao_excitations_2022, kruger_triple_2023}, BaCo$_2$(AsO$_4$)$_2$~\cite{zhong_weak_2020, halloran_geometrical_2023, maksimov_strong_2025},    Na$_3$Ni$_2$BiO$_6$~\cite{shangguan_one_2023, chen_strength_2024, konieczna_reveal_2026}, for which the magnitude or even necessity of the non-Heisenberg exchange components, including the Kitaev interactions, remains debated. By establishing a framework for disentangling competing interactions, our Pareto analysis provides a practical route for determining the minimal model that is necessary to understand unconventional magnetic states in frustrated quantum materials.

\begin{acknowledgments} 
We acknowledge helps from Pascal Manuel with experiments on WISH. We wish to acknowledge helpful discussions with Gang Chen, Yuan Li, Zhentao Wang, Jie Ma, Tom Fennell, and Bruce Normand. Works at the University of Science and Technology of China were supported by National Key R\&D Program of China under the Grant No.~2024YFA1613100 and the National Natural Science Foundation of China (NSFC) under the Grant No. 12374152. Xuefeng Sun acknowledges the support from the National Key R\&D Program of China under the Grant No. 2023YFA1406500 and the NSFC under the Grant Nos. 12574042 and 12274388. Shunhong Zhang acknowledges the support from the NSFC under the Grant No.~12474243. Diffuse neutron scattering experiments on powder samples were carried out at the China Advanced Research Reactor (CARR). Diffuse neutron scattering experiments on single crystals were carried out at the ISIS Spallation Neutron Source under the user program. INS experiments were carried out at the Materials and Life Science Experimental Facility (MLF) of the Japan Proton Accelerator Research Complex (J-PARC) under the user program (Proposal No. 2023B0351). Neutron diffraction experiments on a single crystal were carried out at the Swiss Spallation Neutron Source (SINQ) at the Paul Scherrer Institute (PSI) under the user program (Proposal No. 20241628). 
\end{acknowledgments}

\newpage
\bibliography{GCO_ref}
\newpage

\setlength{\bibsep}{1em}%
\end{document}